\documentclass[aps,twocolumn,prb]{revtex4}
\usepackage{graphicx,amsmath,bm}

\newcommand{\figurewidth}{\columnwidth}

\newcommand{\abv}[1]{{\overline{\left\langle{#1}\right\rangle}}}
\newcommand{\kB}{k_{\rm B}}
\newcommand{\Po}{{P_{\rm o}}}
\newcommand{\C}{{\cal C}}

\begin{document}
\title{
Non-trivial link overlap distribution\\
in three-dimensional Ising spin glasses}
\author{Guy Hed$^1$ and Eytan Domany$^2$}
\affiliation{ $^1$Department of Science Teaching and
$^2$Department of Physics of Complex Systems,\\
Weizmann Institute of Science, Rehovot 76100, Israel}
\date{\today}

\begin{abstract}
We investigate the distributions of the link overlap, $P(Q)$, in
3-dimensional Ising spin glasses.
We use clustering methodology to identify a set of pairs of states from
different Gibbs states, and calculate its contribution to $P(Q)$.
We find that the distribution over this set does not become trivial as
the system size increases.
\end{abstract}
\maketitle

\section{introduction}

While equilibrium properties of infinite range spin glasses
\cite{SK75} are completely understood within the  framework of
replica symmetry breaking (RSB)
\cite{Parisi79,Parisi80,Parisi83,Mezard84}, spin glasses with short
range interactions are the subject of considerable current debate
and controversy. Open  questions address the nature of the low
temperature phases
\cite{Parisi83,Mezard84,droplet,Fisher87,Newman03} and their
theoretical description. Resolution of these issues by experiments
or simulations is hindered by the  extremely long relaxation time
required for equilibration.

The the Edwards-Anderson model is the most widely studied type of
short-range Ising spin glass
\begin{equation}\label{eq:H}
H({\bm s})=\sum_{(i,j)}J_{ij}s_i s_j \;,
\end{equation}
where the sum is over nearest neighbor sites, $(i,j)$,
of a simple (hyper) cubic lattice with periodic boundary conditions,
$s_i=\pm 1$, and the couplings, $J_{ij}$, are independent random variables
taken from a normal distribution with zero average and standard deviation
$J =1$. For the infinite range Sherington Kirkpatrick (SK) model the
sum in Eq. \ref{eq:H} is over all pairs of sites and $J=1/\sqrt{N}$.

The system (in 3 or more dimensions) has a finite critical
temperature. Recent numerical analysis of 3-dimensional Ising spin
glasses (3DISG) yielded\cite{Katzgraber06} $T_c=0.951(9)$, whereas
for the SK model $T_c=1$. The high temperature phase of the model
is a disordered paramagnet. As the temperature decreases below
$T_c$ the system undergoes a transition into a frozen spin-glass
phase.

In the spin glass phase, the microstates are divided into Gibbs
states; each constitutes an ergodic subset of phase space, i.e. a
maximal subspace that the system can span (or visit) as time tends
to infinity. In a finite system phase space consists of one such
state; however, we identify the infinite volume Gibbs states with
subsets of the phase space surrounded by free energy barriers,
whose height diverges as $N \rightarrow \infty$. Here the term
``Gibbs states" refers to such subsets.

Extensive recent numerical investigations, done at finite $T$
\cite{Marinari98,Katzgraber01,SHICS}, as well as ground state
analysis \cite{Marinari00,Franz00b,Houdayer00,HedPRL}, suggest
evidence for a multiplicity of Gibbs states in the low temperature
phase of the 3DISG. The most widely measured properties are
\begin{equation}\label{eqqab}
q({\bm s},{\bm t}) = N^{-1} \sum_{i=1}^N s_i t_i \;,
\end{equation}
the site overlap between any two microstates $\bm s$ and $\bm t$,
the {\it global} distribution of $q$,
\begin{equation}\label{eqpq}
p_J(q) = Z^{-2} \sum_{{\bm s}} e^{-H({\bm s})} \sum_{{\bm
t}} e^{-H({\bm t})} \delta(q({\bm s},{\bm t})-q) \;,
\end{equation}
and, in particular, $p(q) = \overline{p_J(q)}$, the distribution
averaged over the random variables $\{J_{ij}\}$. In (\ref{eqpq})
$Z$ is the partition function of the system.

There is general agreement that in 3DISG the averaged distribution
$p(q)$ is not trivial (that is, does not converge to a
$\delta$-function in the infinite volume limit) and is not self
averaging, in agreement with RSB theory.

As to the droplet theory \cite{droplet,Huse87}, although its common
interpretation involves a trivial $p(q)$ \cite{Fisher86}, it
explicitly predicts only the triviality of the \emph{local} $p(q)$ -
the overlap distribution over a finite (large) window in an infinite
system. That is, \emph{locally} there is only one Gibbs state for
the system (up to a spin-flip symmetry), so that when a finite
window of an infinite system is viewed, the system will be almost
always in this Gibbs state.

In order to test this prediction
numerically, one should observe the site overlap in a finite
constant window as the system size increases \cite{Palassini99}. An
alternative is to measure the link overlap
\begin{equation}\label{eqQab}
Q({\bm s},{\bm t}) = N_{\rm b}^{-1} \sum_{(i,j)} s_i s_j t_i t_j \;,
\end{equation}
summing over all the $N_{\rm b}$ bonds in the system, e.g. over
all nearest neighbors pairs in the case of 3DISG. The distribution
of the link overlap, $P_J(Q)=\langle \delta(Q({\bm s},{\bm t})-Q) \rangle$,
is defined similarly to $p_J(q)$ in Eq. \ref{eqpq}. The average
over realizations is $P(Q)=\overline{P_J(Q)}$.

According to the droplet theory $P(Q)$ is trivial and consequently
$P_J(Q)=P(Q)$. This was predicted earlier from scaling analysis of
numerical results \cite{McMillan84,Bray84}. Newman and Stein
\cite{Newman03,Newman05} showed that triviality of $P(Q)$ for
given boundary conditions may be deduced from general
considerations even if one relaxes some of the scaling
assumptions of the droplet theory. However, according to RSB
$P(Q)$ is not trivial. Unlike the case of $p(q)$, the triviality
of $P(Q)$ is still an open question
\cite{Parisi00,Krzakala00,Newman03,contucci06}.

\section{Numerical simulations}

In the present work we analyze 3DISG systems of sizes $L=8,10,12$.
We generated 65 realizations of the disorder, $\{J_{ij}\}$, for
each system size. For each realization we produced a weighted
sample of microstates. We used simulated tempering Monte Carlo
\cite{MarinariST}.
The number of temperatures, $N_T$, and their
values were determined so that in each inter-temperature
step the transition probability is larger than 0.25.

We used $N_T=16$ for $L=8$, and $N_T\geq27$ for $L=10,12$. The
lowest and highest temperatures were $T_1=0.24J/\kB$ and
$T_{N_T}=2J/\kB$ ($T_{N_T}=0.92J/\kB$ for $L=8$). For every
realization, we took $700 N_T$ samples (i.e.
microstate-temperature pairs). We ran $20{N_T}^2$ sweeps to
thermalize the system, and the same number of sweeps between
samples.

In most realizations the temperatures of consecutive
samples were decorrelated, so that the system often went above
$T_c$, where it lost spin correlations. For some realizations the
decorrelation times were as large as 10 sampling cycles.

\section{Link overlap distribution}

One approach used to study the link overlap is to measure the
fractal dimension of the surface of low energy excitations
\cite{Krzakala00,Katzgraber01}. Such studies are limited by the
small system sizes available for numerical investigation.

\begin{figure}[t]
\includegraphics[width=\figurewidth]{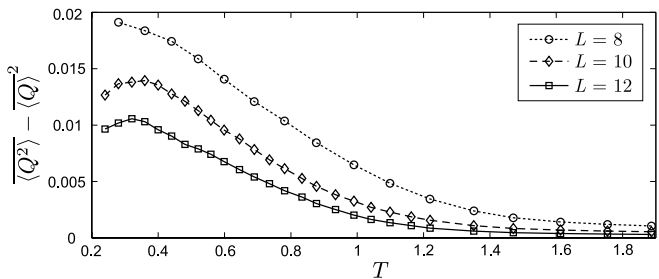}
\caption{The variance of the link overlap distribution, $P(Q)$, for
3DISG systems. The error bars are larger then the values of the
variance.} \label{figvQ}
\end{figure}

Another approach is to calculate directly the moments of the link
overlap distribution, $P(Q)$ \cite{Contucci05,ContucciUM}. In Fig.
\ref{figvQ} we present $\abv{Q^2}-\abv{Q}^2$, the second moment of
$P(Q)$. As the system size increases, the second moment decreases
(apparently towards zero) at all temperatures, indicating that
$P(Q)$ becomes trivial as $L\rightarrow \infty$.

However, characterization of the behavior of a distribution on the
basis of a single parameter, such as a fractal dimension or a
particular moment, may be misleading. This was demonstrated in Ref.
\onlinecite{HedEPL} for the site overlap distribution, and the same
holds for the link overlap distribution.

The distribution $P_J(Q)$, measured for a particular realization
of the random $\{J_{ij}\}$, may consist of many modes, which
correspond to different peaks in the graph of $P_J(Q)$. Each such
mode is related to the distribution of overlaps between two given
Gibbs states. We denote by $G_1$ the Gibbs state with the highest
statistical weight.

As temperature decreases the variation of the statistical weight
of the Gibbs states increases. In particular, the weight of $G_1$
approaches 1 as $T$ goes to zero. Thus, at low temperatures the
most dominant mode of $P_J(Q)$ is
\begin{equation}\label{eqPlab}
P_{1}(Q) = {\rm Prob} (\, Q({\bm s},{\bm t}) = Q \, |\,
{\bm s},{\bm t}\in G_1 \,)
\end{equation}
- the distribution of overlaps within $G_1$ - which is trivial in
the thermodynamic limit. At small system sizes, as the ones
considered in this work, the peak of $P_{1}(Q)$ is still wide, and
it narrows as $L$ increases. Since even at finite (albeit low)
temperatures $P_{1}(Q)$ dominates the distribution, one would
expect a clear decrease of the variance of $P(Q)$ with size, as
observed in Fig. \ref{figvQ}, even if $P(Q)$ is
non-trivial in the thermodynamic limit. Hence decrease of the
second moment of the distribution with size may capture nothing
but the narrowing of the dominant single peak part of
a non-trivial $P(Q)$; in order to test non-triviality, it is much
more effective to measure the size dependence of a quantity which
is independent of the dominant (but uninformative) $P_1(Q)$.

We present $P(Q)$ in Fig. \ref{figPQ}(a) for two temperatures, one
above and one below $T_c$, for three system sizes. Above $T_c$,
$P(Q)$ consists of a single peak which narrows as the $L$
increases;, approaching $\delta(Q-\abv{Q})$. Below $T_c$ the
distribution consists of a dominant peak, whose main contribution
comes from $P_{1}(Q)$, and a tail at lower values of $Q$. As $L$
increases, the peak narrows and since the weight of the peak is much
larger then that of the tail, a single measured parameter (say
moment) may reflect mainly the convergence of the peak to a $\delta$
function, which is expected irrespective of whether $P(Q)$ is
trivial or not. To address this question one has to assess whether
the weight of the tail of $P(Q)$, say, in the range $0 \leq Q \leq
0.8$, decreases with increasing system size.

\begin{figure}[t]
\includegraphics[width=\figurewidth]{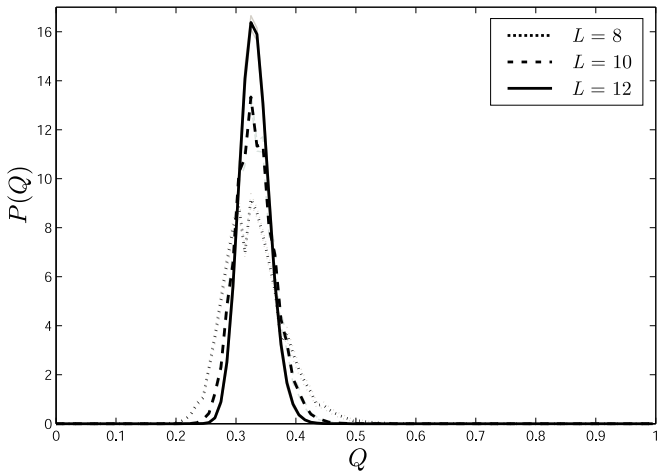}
\caption{(a) The link overlap distribution, $P(Q)$ measured for
3DISG systems at $T=1.35J/\kB >T_c$. The distribution converges to a
$\delta$-peak. The grey lines above and below each curve indicate
the error range.} \label{figPQ}
\end{figure}

\begin{figure}[t]
\includegraphics[width=\figurewidth]{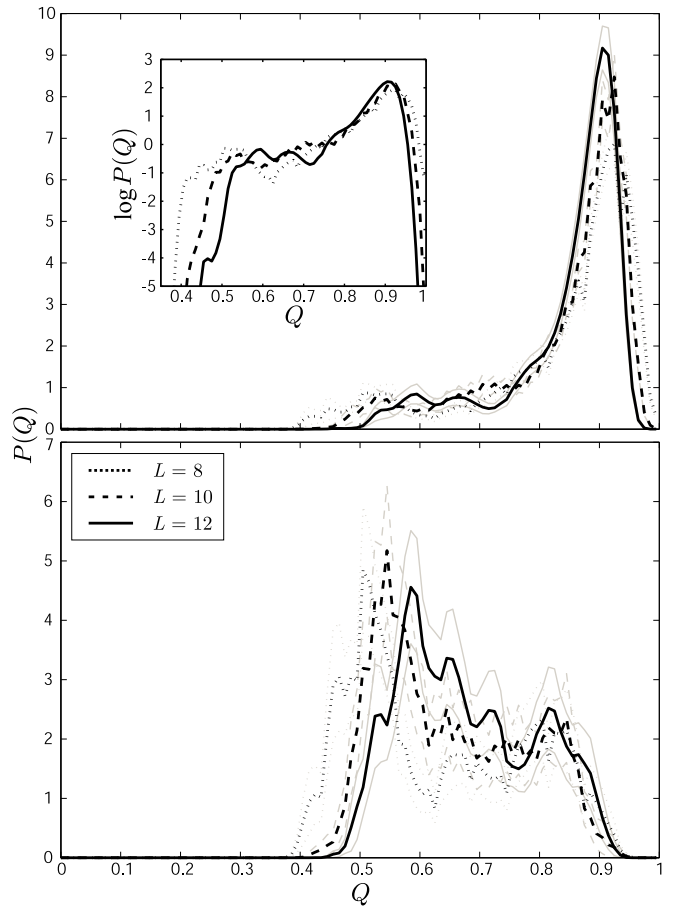}
\caption{(a) The link overlap distribution measured for 3DISG
systems at $T=0.28J/\kB <T_c$, in linear plot and linear-log plot
(inset). (b) The distribution $\Po(Q)$ for $T=0.28J/\kB$, which
gives a finite contribution to $P(Q)$. All distributions are
normalized to unity. The grey lines above and below each curve
indicate the error range.} \label{figPQ}
\end{figure}

If $P(Q)$ is to have a non-trivial component at the
thermodynamic limit, it must originate from pairs of microstates
that belong to different Gibbs states. Ideally, we would like
to calculate the link overlap distribution over all such pairs.
In finite systems this is not possible since we cannot
identify the Gibbs states unambiguously. However, the coarsest
partitions of phase space \emph{can} be  observed clearly, even
for very small systems \cite{HedPRL,SHICS}. This partition
consists of two sets of microstates, $\C_1$ and $\C_2$, which we
identify by the {\it average linkage} agglomerative clustering
algorithm \cite{Jain88}.

The statistical weights of the sets $\C_1$ and $\C_2$
remain finite as the system size increases, and the
average site overlap between them remains different from the
overlap between two microstates that belong to the same set
\cite{SHICS}. Therefore the contribution of the set
of pairs of microstates
$\{ ({\bm s},{\bm t}) \,|\, {\bm s}\in\C_1,{\bm t}\in\C_2 \}$
to $P(Q)$ remains finite as the system size increases.

Following the method described in Ref. \onlinecite{HedEPL}
for $p(q)$, we define
\begin{equation}\label{eqPlo}
\Po(Q)={\rm Prob} (\, Q({\bm s},{\bm t}) = Q \,|\,
{\bm s}\in\C_1,{\bm t}\in\C_2 \,) \;.
\end{equation}
In order to move from a state in $\C_1$ to a state in $\C_2$,
some groups of spins (domains) have to be flipped.  We refer to
the surfaces that separate these flipped domains from spins that
did not flip as domain walls. If these domain walls have a
vanishing density, the fraction of links affected by switching one
micro-state from $\C_1$ to $\C_2$ goes to zero as the system size
increases, and hence at the thermodynamic limit we will have
$\Po(Q)=P(Q)$.

In Fig. \ref{figPQ}(a) we observe that for 3DISG at $T=0.28J/\kB$,
$P(Q)$ seems to converge to a peak at $Q\simeq 0.9$. If $\Po(Q)$
does not converge to the same distribution, we expect it to have a
non-vanishing weight at $Q<0.9$. We calculate $\int_0^{0.8}\Po(Q)
dQ$. For $L=8,10,12$ the values of the integral are 0.81(5),
0.83(4), 0.80(4) respectively, suggesting that the weight of
$\Po(Q)$ for $0 \leq Q \leq 0.8$ does not decrease. Thus, $\Po(Q)$
does not converge to the peak at $Q\simeq 0.9$ which dominates
$P(Q)$.

These results imply that
the domain walls between $\C_1$ and $\C_2$ occupy a finite fraction
of the volume of the system. Since $\C_1$ and $\C_2$
remain finite and distinguishable as the system size increases
\cite{SHICS}, $\Po(Q)$ (shown in Fig. \ref{figPQ}(b)) constitutes
a finite contribution to $P(Q)$. Consequently, $P(Q)$ is not
trivial at the thermodynamic limit.

\section{discussion}

According to RSB theory, $P(Q)$ is not trivial. More over, all
overlap measures are equivalent \cite{Parisi00,Contucci03}, so $Q$
is determined by the value of the site overlap $q$. This statement
is trivial for the SK model, where the sum in  Eq. \ref{eqQab} is
over all pairs of sites, so
$Q({\bm s},{\bm t})=(q^2({\bm s},{\bm t}) - N^{-1})/(1-N^{-1})$.
In 3DISG the site overlap and bond overlap
have different roles: given two states of the system, the site
overlap is given by the volume of spin-domains flipped between the
states, and the link overlap by the surface area of the domain
walls.

Recent numerical investigation of 3DISG systems \cite{contucci06}
indicated that the two overlaps are equivalent. This result along with
previous numerical evidence for the non-triviality of the site overlap,
$p(q)$, yield a non-trivial $P(Q)$ and support the results presented here.

\begin{acknowledgments}
The authors acknowledge useful discussions with Ido Kanter, and a
most helpful correspondence with A. Peter Young, Pierluigi Contucci
and Cristian Giardin\`a.
\end{acknowledgments}

\ \\ \  \\ \  \\ \  \\



\end{document}